\documentclass[runningheads]{llncs}
\usepackage{graphicx}
\usepackage{amsmath}
\usepackage{amssymb}
\usepackage{booktabs}
\usepackage[acronym,xindy,toc]{glossaries}
\usepackage{bm} % bold math symbols
\usepackage{tabularx} % tables
\usepackage{makecell} % formatting table cells
\usepackage{mathtools}
\usepackage{multirow}
\usepackage[export]{adjustbox}

\usepackage{hyperref}
\hypersetup{pagebackref,breaklinks,colorlinks}

\newcommand{\interval}[1]{{\scriptsize\textpm#1}}
\newcommand{\pred}[1]{{#1}^\mathrm{p}}
\newcommand{\gt}[1]{{#1}^\mathrm{gt}}
\newcommand{\Mpred}{\pred{\mathcal{M}}}
\newcommand{\Mgt}{\gt{\mathcal{M}}}
\newcommand{\Ppred}{\pred{\mathcal{P}}}
\newcommand{\Pgt}{\gt{\mathcal{P}}}
\newcommand{\stimes}{{\times}} % No space between axb

\DeclarePairedDelimiter\abs{\lvert}{\rvert}%
\newcommand{\expnumber}[2]{{#1}\mathrm{e}{#2}} % exponential notation

\usepackage{comment}
\usepackage[capitalize]{cleveref}
\crefname{subsection}{Sec.}{Secs.}
\Crefname{subsection}{subsection}{Sections}
\Crefname{table}{Table}{Tables}
\crefname{table}{Tab.}{Tabs.}

\newacronym[shortplural={D$_{\text{dye}}$}, longplural={donor dye, ex. Alexa 488}]{ddye}{D$_{\text{dye}}$}{donor dye, ex. Alexa 488}

\newacronym[description={\glslink{r0}{F\"{o}rster distance}}]{R0}{$R_{0}$}{F\"{o}rster distance}

\newglossaryentry{r0}
{
  name=\glslink{R0}{\ensuremath{R_{0}}},
  text=F\"{o}rster distance,
  description={F\"{o}rster distance, where 50\% ...}, 
  sort=R
}

\newglossaryentry{kdeac}
{
  name=\glslink{R0}{\ensuremath{k_{DEAC}}},
  text=$k_{DEAC}$, 
  description={is the rate of deactivation from ... and emission)}, 
  sort=k
}

\newacronym[shortplural={TUM}, longplural={Technical University of Munich}]{tuma}{TUM}
{Technical University of Munich}

\newglossaryentry{tum}
{
  name = TUM,
  description = {A university in Germany, Bavaria, in the city of Munich},
  plural = TUM
}

\newglossaryentry{computer}
{
  name=computer,
  description={is a programmable machine that receives input,
               stores and manipulates data, and provides
               output in a useful format}
}

\newglossaryentry{nonlin}
{
	name={\glslink{nonlin}\ensuremath{h}},
	description={A nonlinearity}
}

\newacronym[shortplural={NNs}, longplural={neural networks}]{nn}{NN}{neural network}
\newacronym[shortplural={GNNs}, longplural={graph neural networks}]{gnn}{GNN}{graph neural network}
\newacronym[shortplural={CNNs}, longplural={convolutional neural networks}]{cnn}{CNN}{convolutional neural network}
\newacronym{ai}{AI}{artificial intelligence}
\newacronym{dl}{DL}{deep learning}
\newacronym{iou}{IoU}{Intersection over Union}
\newacronym{asp}{ASP}{Anatomic Segmentation using Proximity}
\newacronym{clasp}{CLASP}{Contrained Laplacian-based ASP}
\newacronym{csf}{CSF}{cerebrospinal fluid}
\newacronym{mri}{MRI}{magnetic resonance imaging}
\newacronym{pve}{PVE}{partial volume effect}
\newacronym{sdf}{SDF}{signed distance function}
\newacronym{lns}{LNS}{learned neighborhood sampling}
\newacronym{ct}{CT}{computer tomography}
\newacronym{hd}{HD}{90-percentile Hausdorff distance}
\newacronym{assd}{ASSD}{average symmetric surface distance}
\newacronym{sgd}{SGD}{stochastic gradient descent}
\newacronym{relu}{ReLU}{rectified linear unit}
\newacronym{exprecs}{EXPRECS}{\textbf{EXP}licit \textbf{RE}construction of \textbf{C}ortical \textbf{S}urfaces}
\newacronym{trt}{TRT}{test-retest dataset}
\newacronym{icp}{ICP}{iterative closest-point algorithm}
\newacronym{fcnn}{F-CNN}{fully-convolutional neural network}
\newacronym{mlp}{MLP}{multi-layer perceptron}
\newacronym{mr}{MR}{magnetic resonance}

\newglossaryentry{X}{name={\ensuremath{\mathcal{X}}},description={Input domain of a neural network.}}
\newglossaryentry{Y}{name={\ensuremath{\mathcal{Y}}},description={Output domain of a neural network.}}
\newglossaryentry{L}{name={\ensuremath{\mathcal{L}}},description={Loss function.}}
\newglossaryentry{theta}{name={\ensuremath{\theta}},description={Network parameters.}}
\newglossaryentry{dCD}{name={\ensuremath{d_\mathrm{CD}}},description={Chamfer distance}}
\newglossaryentry{LCD}{name={\ensuremath{\mathcal{L}_\mathrm{CD}}},description={Chamfer loss}}
\newglossaryentry{LCDcurv}{name={\ensuremath{\mathcal{L}_{\mathrm{C}}}},description={Chamfer loss}}
\newglossaryentry{LC-ag}{name={\ensuremath{\mathcal{L}_{\mathrm{rec}}}},description={Class-agnostic Chamfer loss}}
\newglossaryentry{Jacc}{name={\ensuremath{J}},description={Jaccard index}}
\newglossaryentry{dJacc}{name={\ensuremath{d_{\mathrm{Jacc}}}},description={Jaccard distance}}
\newglossaryentry{Dice}{name={\ensuremath{DSC}},description={Dice coefficient}}
\newglossaryentry{Hausd}{name={\ensuremath{d_\mathrm{H}}},description={Hausdorff distance}}
\newglossaryentry{dASSD}{name={\ensuremath{d_\mathrm{AD}}},description={Average symmetric surface distance}}
\newglossaryentry{LCE}{name={\ensuremath{\mathcal{L}_\mathrm{BCE}}},description={Cross entropy loss}}
\newglossaryentry{LNC}{name={\ensuremath{\mathcal{L}_{\mathrm{n}, \, intra}}},description={Normal consistency loss}}
\newglossaryentry{Lcos}{name={\ensuremath{\mathcal{L}_{\mathrm{n}, \, inter}}},description={Cosine loss}}
\newglossaryentry{LLapabs}{name={\ensuremath{\mathcal{L}_{\mathrm{Lap}, \, abs}}},description={Laplacian smoothing loss w.r.t. absolute vertex coordinates}}
\newglossaryentry{LLaprel}{name={\ensuremath{\mathcal{L}_{\mathrm{Lap}, \, rel}}},description={Laplacian smoothing loss w.r.t. relative vertex coordindates}}
\newglossaryentry{LLap}{name={\ensuremath{\mathcal{L}_{\mathrm{Lap}}}},description={Laplacian smoothing loss}}
\newglossaryentry{Ledge}{name={\ensuremath{\mathcal{L}_\mathrm{edge}}},description={Edge loss}}
\newglossaryentry{phi}{name={\ensuremath{\bm{\phi}}},description={Feature vector}}
\newglossaryentry{diff_co}{name={\ensuremath{\bm{\xi}}},description={Differential coordinates}}
\newglossaryentry{curv}{name={\ensuremath{\bar{\kappa}}},description={Discrete mean curvature}}
\newglossaryentry{Diff_co}{name={\ensuremath{\bm{\Xi}}},description={Matrix of differential coordinates}}
\newglossaryentry{Lap}{name={\ensuremath{\bm{\mathrm{L}}}},description={Laplace operator}}
\newglossaryentry{Lap-u}{name={\ensuremath{\bm{\mathrm{L}}_\mathrm{u}}},description={Laplace operator with uniform weights}}
\newglossaryentry{Lap-cot}{name={\ensuremath{\bm{\mathrm{L}}_\mathrm{cot}}},description={Laplace operator with cotangent weights}}
\newglossaryentry{A}{name={\ensuremath{\mathrm{\textbf{A}}}},description={Adjacency matrix}}
\newglossaryentry{T}{name={\ensuremath{\mathcal{T}}},description={Template mesh}}
\newglossaryentry{D}{name={\ensuremath{\mathrm{\textbf{D}}}},description={Degree matrix}}
\newglossaryentry{n}{name={\ensuremath{n}},description={Number of mesh vertices}}
\newglossaryentry{deg}{name={\ensuremath{\mathrm{deg}}},description={Degree of a vertex}}
\newglossaryentry{N}{name={\ensuremath{\mathcal{N}}},description={Neighborhood of a vertex}}
\newglossaryentry{N_img}{name={\ensuremath{N}},description={The number of pixels or voxels in an image}}
\newglossaryentry{H_img}{name={\ensuremath{H}},description={The height of an image}}
\newglossaryentry{W_img}{name={\ensuremath{W}},description={The width of an image}}
\newglossaryentry{D_img}{name={\ensuremath{D}},description={The depth of an image}}
\newglossaryentry{V}{name={\ensuremath{\mathcal{V}}},description={Set of vertices}}
\newglossaryentry{Vmat}{name={\ensuremath{\bm{\mathrm{V}}}},description={Vertices stacked into a matrix}}
\newglossaryentry{Fmat}{name={\ensuremath{\bm{\mathrm{F}}}},description={Faces stacked into a matrix}}
\newglossaryentry{E}{name={\ensuremath{\mathcal{E}}},description={Set of edges}}
\newglossaryentry{F}{name={\ensuremath{\mathcal{F}}},description={Set of faces}}
\newglossaryentry{lr}{name={\ensuremath{\lambda}},description={Learning rate}}
\newglossaryentry{Lvox}{name={\ensuremath{\mathcal{L}_\mathrm{vox}}},description={Voxel-based loss}}
\newglossaryentry{Lmesh}{name={\ensuremath{\mathcal{L}_\mathrm{mesh}}},description={Mesh-based loss}}
\newglossaryentry{Lmesh_cons}{name={\ensuremath{\mathcal{L}_{\mathrm{mesh},\, cons}}},description={Geometry-consistency loss}}
\newglossaryentry{Lmesh_reg}{name={\ensuremath{\mathcal{L}_{\mathrm{mesh}, \, reg}}},description={Mesh-regularization loss}}
\newglossaryentry{S}{name={\ensuremath{S}},description={The number of mesh-deformation stages}}
\newglossaryentry{n_seg}{name={\ensuremath{L}},description={The number of segmentation outputs (voxel-based)}}
\newglossaryentry{n_struc}{name={\ensuremath{C}},description={The number of surfaces or connected components in one mesh}}
\newglossaryentry{Vrel}{name={\ensuremath{\bm{\Delta}}},description={Matrix of relative coordinates}}
\newglossaryentry{mlw}{name={\ensuremath{\lambda}},description={Mesh-loss weight}}

\usepackage[export]{adjustbox}
\usepackage{array}
\usepackage{arydshln}
\usepackage{booktabs}
\newcommand{\ra}[1]{\renewcommand{\arraystretch}{#1}}
\newcommand\blfootnote[1]{%
  \begingroup
  \renewcommand\thefootnote{}\footnote{#1}%
  \addtocounter{footnote}{-1}%
  \endgroup
}
\begin{document}
%
%\title{Contribution Title\thanks{Supported by organization x.}}
\title{Joint Reconstruction and Parcellation of Cortical Surfaces}
% Joint Surface Reconstruction and Parcellation of the Cerebral Cortex
% Mesh-based parcellation of the cerebral cortex
% Mesh-based  reconstruction and parcellation of the cerebral cortex
% Joint reconstruction and parcellation of cortical surfaces (fromm brain MRI scans)
% 

%
\titlerunning{Joint Reconstruction and Parcellation of Cortical Surfaces}
% If the paper title is too long for the running head, you can set
% an abbreviated paper title here
%
\author{Anne-Marie Rickmann\inst{1,2}$^{*}$ \and
Fabian Bongratz\inst{2}$^*$ \and
Sebastian P\"{o}lsterl\inst{1} \and
Ignacio Sarasua \inst{1,2} \and
Christian Wachinger \inst{1,2}
}
%index{Rickmann, Anne-Marie}
%index{Bongratz, Fabian}
%index{P\"{o}lsterl, Sebastian}
%index{Sarasua, Ignacio}
%index{Wachinger, Christian}

%
\authorrunning{Rickmann and Bongratz et al.}
% First names are abbreviated in the running head.
% If there are more than two authors, 'et al.' is used.
%
\institute{Ludwig-Maximilians-University, Munich, Germany \and
Technical University of Munich, Germany \newline
Lab for Artificial Intelligence in Medical Imaging}

%\institute{anonymous institute}
%
\maketitle              % typeset the header of the contribution
\blfootnote{$^*$ Equal contribution}

\begin{abstract}
The reconstruction of cerebral cortex surfaces from brain MRI scans is instrumental for the analysis of brain morphology and the detection of cortical thinning in neurodegenerative diseases like Alz\-hei\-mer's disease (AD). 
Moreover, for a fine-grained analysis of atrophy patterns, the parcellation of the cortical surfaces into individual brain regions is required. 
%In addition, the parcellation of the cortical surface provides more detailed information about different regions of the cortex. This is in particular useful for the computation of region-based biomarkers of the brain.
For the former task, powerful deep learning approaches, which provide highly accurate brain surfaces of tissue boundaries from input MRI scans in seconds, have recently been proposed. However, these methods do not come with the ability to provide a parcellation of the reconstructed surfaces.
Instead, separate brain-parcellation methods have been developed, which typically consider the cortical surfaces as given, often computed beforehand with FreeSurfer.
%Each of these approaches focuses on one part, either surface reconstruction, voxel based segmentation or surface based parcellation.
In this work, we propose two options, one based on a graph classification branch and another based on a novel generic 3D reconstruction loss, to augment template-deformation algorithms such that the surface meshes directly come with an atlas-based brain parcellation. By combining both options with two of the latest cortical surface reconstruction algorithms, we attain highly accurate parcellations with a Dice score of 90.2 (graph classification branch) and 90.4 (novel reconstruction loss) together with state-of-the-art surfaces.

% Both options  By adding a graph classification branch to the surface reconstruction, we achieve an average surface parcellation Dice score of 90.2. As an alternative, we introduce a novel and generic 3D class-based reconstruction loss to propagate a parcellation on an input atlas to the predicted surfaces, which yields an average Dice of 90.4. Both of these algorithmic extensions largely maintain the accuracy of the surface reconstructions and, in addition, yield highly accurate brain parcellations in seconds.

%that reconstruct cortical surfaces using a combination of graph convolutional and convolutional neural networks, with a surface based parcellation approach. 
%We show that adding classification layers to the graph convolutional network yields highly accurate surface parcellation and even improves the quality of the surface recontruction.
%The abstract should briefly summarize the contents of the paper in
%150--250 words.

%\keywords{First keyword  \and Second keyword \and Another keyword.}
\end{abstract}

\section{Introduction}
The reconstruction of cerebral cortex surfaces from brain MRI scans remains an important task for the analysis of brain morphology and the detection of cortical thinning in neurodegenerative diseases like Alzheimer's disease (AD)~\cite{Roe2021}. Moreover, an accurate parcellation of the cortex into distinct regions is essential to understand its inner working principles as it facilitates the location and the comparison of measurements~\cite{glasser2016multi,vanEssen2012ParcellationsAH}.
%provides valuable information about cortical subregions~\cite{glasser2016multi}.
While voxel-based segmentations are useful for volumetric measurements of subcortical structures, they are merely suited to represent the tightly folded and thin (thickness in the range of few millimeters~\cite{mai2011human}) geometry of the cerebral cortex.
%In particular, the voxel-based parcellation of the cortex is not optimal for measurements like cortical thickness, which need to be highly accurate (e.g., differences in cortical thickness between AD patients and healthy controls are usually in sub-millimeter range).

The traditional software pipeline FreeSurfer~\cite{fischl2012freesurfer}, which is commonly used in brain research, addresses this issue by offering a surface-based analysis in addition to the voxel-based image processing stream. More precisely, the voxel stream provides a voxel-based segmentation of the cortex and subcortical structures, whereas the surface-based stream creates cortical surfaces and a cortex parcellation on the vertex level. To this end, FreeSurfer registers the surfaces to a spherical atlas. Cortical thickness can be computed from these surfaces with sub-millimeter accuracy and different regions of the brain can easily be analyzed given the cortex parcellation. Yet, the applicability of FreeSurfer is limited by its lengthy runtime (multiple hours per brain scan).

Recently, significantly faster deep learning-based approaches for cortical surface reconstruction have been proposed~\cite{Bongratz_2022_CVPR,santacruz2021,lebrat2021corticalflow,ma2021b}; they reconstruct cortical surfaces from an MRI scan within seconds. To date, however, these methods do not come with the ability to provide a parcellation of the surfaces. At the same time, recent parcellation methods~\cite{cucurull2018convolutional,gopinath2019graph} usually rely on FreeSurfer for the extraction of the surface meshes. A notable exception is~\cite{gopinath2021}, which, however, is not competitive in terms of surface accuracy. 

\begin{figure}[t]
 \centering
 \includegraphics[keepaspectratio,width=\textwidth,height=\textheight]{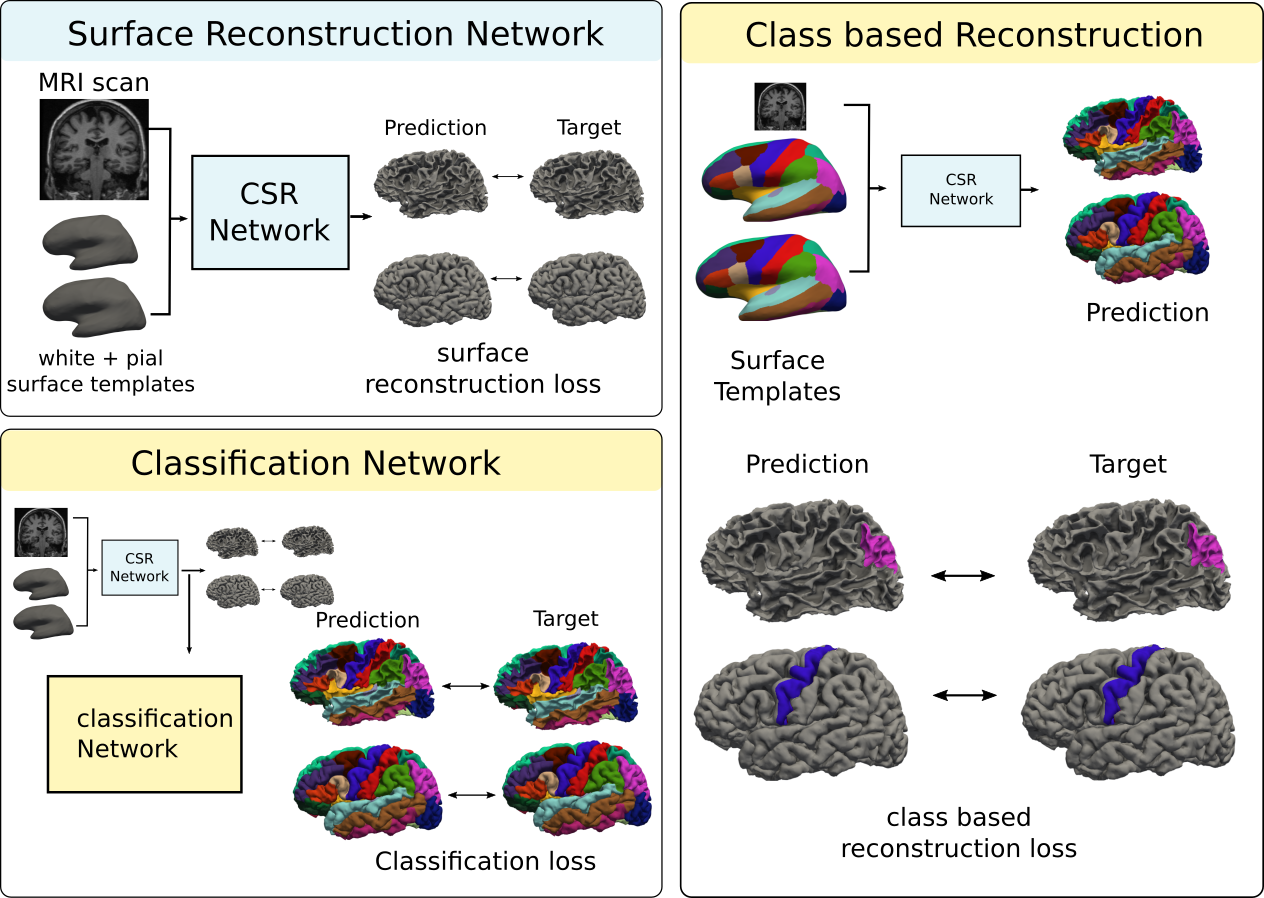}
 \caption{Overview of surface reconstruction networks with our extensions for learning cortex parcellation. Bottom left: A classification network is added after the deformation network and trained with a classification loss on the vertex-wise class predictions. Right: The deformation network takes surface templates with parcellation labels (from a population atlas) as input and the reconstruction loss is computed separately for each class.}
 \label{fig:overview}
\end{figure}

In this work, we close this gap by augmenting two state-of-the-art cortical surface reconstruction (CSR) methods~\cite{Bongratz_2022_CVPR,lebrat2021corticalflow} with two different parcellation approaches in an end-to-end trainable manner. Namely, we extend the CSR networks with a graph classification network and, as an alternative, we propagate template parcellation labels through the CSR network via a novel class-based reconstruction loss. Both approaches are illustrated in Figure~\ref{fig:overview}.
We demonstrate that both approaches yield highly accurate cortex parcellations on top of state-of-the-art boundary surfaces. 

\section{Related Work}
In the following, we will briefly review previous work related to corical surface reconstruction and cortex parcellation. While we focus on joint reconstruction and parcellation, the majority of existing methods solves only one of these two tasks at a time, i.e., cortex parcellation \emph{or} cortical surface reconstruction.

Convolutional neural networks (CNNs) remain a popular choice for medical image segmentation and they have been applied successfully to the task of cortex parcellation. For example, FastSurfer~\cite{henschel2020} replaces FreeSurfer’s voxel-based stream by a multi-view 2D CNN. Similar approaches~\cite{HUO2019105,coupe2020assemblynet} have been proposed based on 3D patch-based networks. However, the computation of cortical biomarkers based on fully-convolutional segmentations is ultimately restricted by the image resolution of the input MRI scans and the combination with the FreeSurfer surface stream is not efficient in terms of inference time.

Deep learning-based parcellation methods operating on given surface meshes (typically pre-computed with FreeSurfer) have also been presented in the past. For example, the authors of~\cite{cucurull2018convolutional} investigate different network architectures for the segmentation of two brain areas. They found that graph convolution-based approaches are more suited compared to multi-layer perceptrons (MLPs). Similarly, the method presented in~\cite{eschenburg2021learning} parcellates the whole cortex using graph attention networks. In contrast, the authors of~\cite{gopinath2019graph} utilize spherical graph convolutions, which they find to be more effective than graph convolutions in the Euclidean domain. All of these vertex classifiers consider the surface mesh as given.

To avoid the lengthy runtime of FreeSurfer for surface generation, deep learning-based surface reconstruction approaches focus on the fast and accurate generation of cortical surfaces from MRI. These approaches can be grouped into implicit methods~\cite{santacruz2021}, which learn signed distance functions (SDFs) to the white-to-gray-matter and gray-matter-to-pial interfaces, and explicit methods~\cite{Bongratz_2022_CVPR,lebrat2021corticalflow}, which directly predict a mesh representation of the surfaces. The disadvantage of implicit surface representations is the need for intricate mesh extraction, e.g., with marching cubes~\cite{lewiner2003efficient}, and topology correction. This kind of post-processing is time-consuming and can introduce anatomical errors~\cite{fischl2012freesurfer}. In contrast, Vox2Cortex~\cite{Bongratz_2022_CVPR} and CorticalFlow~\cite{lebrat2021corticalflow} deform a template mesh based on geometric deep learning. More precisely, Vox2Cortex implements a combination of convolutional and graph-convolutional neural networks for the template deformation, whereas CorticalFlow relies on the numerical integration of a deformation field predicted in the image domain. Both of these approaches provide highly accurate cortical surfaces without the need for post-processing.

To the best of our knowledge, SegRecon~\cite{gopinath2021} is the only approach that simultaneously learns to generate cortical surfaces and a dedicated parcellation. The authors trained a 3D U-Net to learn a voxel-based SDF of the white-to-gray-matter interface and spherical coordinates in an atlas space. After mesh extraction and time-intense topology correction, the atlas parcellation can be mapped to the surfaces. Although this method can extract a white matter surface from an MRI (reported Hausdorff distance 1.3 mm), the focus of SegRecon lies on the parcellation and the respective surface reconstructions are not competitive with recent algorithms specifically designed for this purpose. 
In contrast to SegRecon, we leverage explicit cortex reconstruction methods since they have shown to yield more accurate surfaces compared to their implicit counterparts~\cite{Bongratz_2022_CVPR,lebrat2021corticalflow}. 

%We demonstrate two different ways to extend these approaches for the task of cortex parcellation on a per-vertex level: one based on a graph neural network (GNN) classifier and another based on template matching via a class-based reconstruction loss. This leads to \emph{end-to-end} trainable networks that solve two tasks at a time, cortex reconstruction and parcellation, while maintaining all advantages of explicit reconstruction methods. Namely, the main advantage is that convenient and accurate surface meshes are obtained without further post-processing. 

%and another  to perform an explict surface reconstruction since the latest research  has demonstrated that learning explicit surfaces is more appropriate, and that graph convolutions are suitable to learn a surface-based parcellation. Therefore, we combine two state-of-the-art explicit surface reconstruction methods~\cite{Bongratz_2022_CVPR,lebrat2021corticalflow} with graph convolutional classification layers. We further investigate how this multi-task approach benefits the individual subtasks.

\section{Method} \label{sec:method}
We build upon very recent work in the field of cortical surface reconstruction and propose to extend the respective methods to endow the reconstructed surfaces with a jointly learned parcellation. In particular, we base our work on Vox2Cortex~\cite{Bongratz_2022_CVPR} and CorticalFlow~\cite{lebrat2021corticalflow}, two mesh-deformation methods that have shown state-of-the-art results for the extraction of cortical surfaces. Both of these methods take a 3D  MRI scan and a mesh template as input and compute four cortical surfaces simultaneously, the white-matter and the pial surfaces of each hemisphere. 

\subsection{Surface Reconstruction Methods}
\subsubsection{Vox2Cortex:}
Inspired by previous related methods~\cite{kong2021b,wang2018,wickramasinghe2020}, Vox2Cortex (V2C)~\cite{Bongratz_2022_CVPR} consists of two neural sub-networks, a CNN that operates on voxels and a GNN responsible for mesh deformation. Both sub-networks are connected via feature-sampling modules that map features extracted by the CNN to vertex locations of the meshes. To avoid self-intersections in the final meshes, which is a common problem in explicit surface reconstruction methods, Vox2Cortex relies on multiple regularization terms in the loss function. The deformation of the template mesh is done in a sequential manner, i.e., multiple subsequent deformation steps that build upon each other lead to the final mesh prediction.

\subsubsection{CorticalFlow:}
In contrast to Vox2Cortex, which predicts the mesh-deformation field on a per-vertex basis, CorticalFlow (CF)~\cite{lebrat2021corticalflow} relies on a deformation field in image space. To map it onto the mesh vertices, CorticalFlow interpolates the deformation field at the respective locations. Similarly to Vox2Cortex, the deformation is done step-by-step and each sub-deformation is predicted by a 3D UNet. To avoid self-intersections, the authors propose an Euler integration scheme of the flow fields. The intuition behind using a numerical integration is that by choosing a sufficiently small step size $h$, the mesh deformation is guaranteed to be diffeomorphic and, thus, intersection-free. However, this guarantee does not hold in practice due to the discretization of the surfaces~\cite{lebrat2021corticalflow}. In our experiments, we apply only a single integration step to reduce training time and memory consumption (also at training time). 

\subsection{Surface Parcellation}
For the parcellation of the human cortex, there exist multiple atlases based on, e.g., structural or functional properties according to which different brain regions can be distinguished. Commonly used atlases are the Desikan-Killiany-Tourville (DKT)~\cite{klein2012,desikan2006} or Destrieux~\cite{destrieux2010} atlas, which are both available in FreeSurfer. 
We use FreeSurfer surfaces as pseudo-ground-truth meshes with parcellation labels from the DKT atlas and smoothed versions of the FreeSurfer fsaverage template as mesh input. 

\subsubsection{Classification network:}
Previous work~\cite{cucurull2018convolutional,eschenburg2021learning} has shown that GNN-based classification networks can provide accurate cortex parcellations.
Therefore, we extend the CSR networks with a classification branch consisting of three residual GNN blocks, each with three GNN layers. 
We hand the predicted mesh with vertex features (extracted by the Vox2Cortex GNN) or just vertex coordinates (from CorticalFlow) as input to the GNN classifier. As output, we obtain a vector of class probabilities for each vertex. After a softmax layer, we compute a cross-entropy loss between the predicted classes and ground-truth classes of the closest points in the target mesh. 
In combination with Vox2Cortex, we integrate the classification network after the last mesh-deformation step and train the CSR and classification networks end-to-end.
In combination with CorticalFlow, we also add the classification layer after the last deformation and freeze the parameters of the U-Nets of the previous steps. In our experiments, we found that adding the classification network in each of the iterative optimization steps leads to training instability, hence we only add it in the last iteration.

\subsubsection{Class-based reconstruction:}
As both Vox2Cortex and CorticalFlow are template-deformation approaches, we propose to propagate the atlas labels of the DKT atlas through the deformation process. More precisely, we enforce the respective regions from the template to fit the labeled regions of the FreeSurfer meshes by using a modified \emph{class-based} reconstruction loss. This loss function is agnostic to the concrete implementation of the reconstruction loss, e.g., it can be given by a Chamfer distance as in CorticalFlow or a combination of point-weighted Chamfer and normal distance as in Vox2Cortex. Let $\Ppred$ and $\Pgt$ be predicted and ground-truth point sets sampled from the meshes $\Mpred$ and $\Mgt$, potentially associated with normals. Further, let $\mathcal{L}_{\mathrm{rec}}(\Ppred_c, \Pgt_c)$ be any reconstruction loss between the point clouds of a certain parcellation class $c \in C$. Then, we compute the class-based reconstruction loss as
\begin{equation} \label{eq:class-based Chamfer distance}
\textstyle
	%\begin{split}	
    \mathcal{L}_{\mathrm{rec,}class}(\Ppred, \Pgt) = \frac{1}{\abs{C}} \sum_{c \in C} \mathcal{L}_{\mathrm{rec}}(\Ppred_c, \Pgt_c) .
    %\end{split}
\end{equation}
Intuitively, the predicted points of a certain atlas class ``see'' only the ground-truth points of the same class. We depict this intuition also in \Cref{fig:overview}B. By construction of this loss, the parcellation of the deformed template and the ground-truth parcellation are aligned. Compared to the classification network, this approach has the advantage that ``islands'', i.e., small wrongly classified regions, cannot occur on smooth reconstructed meshes.

%\section{Experiments}
%\begin{itemize}
%    \item comparison with sota voxel based parcellation methods: nnUnet?
%    \item comparison with Freesurfer and Fastsurfer
%    \item Hypothethis: Mesh Quality: Vox2Cortex with and without parcellation --> with parcellation increases the quality of meshes
%    \item mesh quality: compare with Cortical Flow
%    \item Group analysis? -- per parcel thickness measures - comparison to freesurfer
%    \item Ablation study? - not sure yet because the method has not been developed yet
%    \item parcellation accuracy: Use Mindboggle dataset (contains manual labels) - compare to Freesurfer (slightly biased towards freesurfer as Mindboggle labels are corrected Freesurfer labels) and spectral graph conv paper (Herve) 

\subsection{Experimental Setup} \label{sec:experimental_setup}

\textbf{Data:}
We train our models on 292 subjects of the publicly available  OASIS-1 dataset~\cite{oasis} and use 44 and 80 subjects for validation and testing, respectively. Overall, 100 subjects in OASIS-1 have been diagnosed with very mild to moderate Alzheimer's disease. We based our splits on diagnosis, age, and sex in order to avoid training bias against any of these groups.
%\textbf{ADNI}
%As a second test dataset, we use a subset of 323 subjects of the Alzheimer's Disease Neuroimaging Initiative (ADNI) (\url{http://adni.loni.usc.edu}). This dataset provides MRI T1 scans for subjects with Alzheimer's Disease, Mild Cognitive Impairment, and healthy subjects, of which we have respectively 50, 149, 124 in our subset. We only use baseline scans of the longitudinal study in this work.

\noindent
\textbf{Pre-processing:}
We use the FreeSurfer software pipeline, version v7.2~\footnote[1]{available at \url{https://surfer.nmr.mgh.harvard.edu/}}, as silver standard for training and evaluation of our models. More precisely, we follow the setup in~\cite{Bongratz_2022_CVPR,lebrat2021corticalflow} and use the orig.mgz files and white and pial surfaces generated by FreeSurfer and register the MRI scans to the MNI152 space (rigid and subsequent affine registration). Further, we subsampled the surface meshes to about 40,000 vertices per surface using quadric edge collapse decimation~\cite{garland1997}. Images are padded to have shape $192 \stimes 208 \stimes 192$. For Vox2Cortex experiments, we resize the images after padding to $128 \stimes 144 \stimes 128$ voxels as done in the original paper~\cite{Bongratz_2022_CVPR}. We use min-max-normalization of intensity values to scale to $[0, 1]$.

\noindent
\textbf{Training:}
For the computation of the reconstruction losses, we sample 50,000 points from the predicted and ground-truth meshes in a differentiable manner~\cite{gkioxari2019}. We interpolate curvature information of a sampled point using the barycentric weights from the respective triangle vertices and assign the point class of the closest vertex to a sampled point. For training CorticalFlow, we use an iterative procedure as described by~\cite{lebrat2021corticalflow}, i.e., freezing the UNet(s) of steps 1 to $i-1$ when training UNet $i$. We further use the AdamW optimizer~\cite{loshchilov2018decoupled} with weight decay $\expnumber{1}{-4}$ and a cyclic learning rate schedule~\cite{smith2017cyclic} for the optimization of the networks. As input to the deformation networks, we leverage the fsaverage templates in FreeSurfer and smooth them extensively using the HC Laplacian smoothing operator implemented in MeshLab~\cite{cignoni2008meshlab}. We provide a list of all model parameters, which we adopt from the Vox2Cortex and CorticalFlow papers, in the supplemental material. Our implementation is based on PyTorch~\cite{pytorch} and PyTorch3d~\cite{ravi2020} and we trained on an Nvidia Titan RTX GPU.

\begin{figure}[t]
 \centering
 \includegraphics[keepaspectratio,width=\textwidth,height=\textheight]{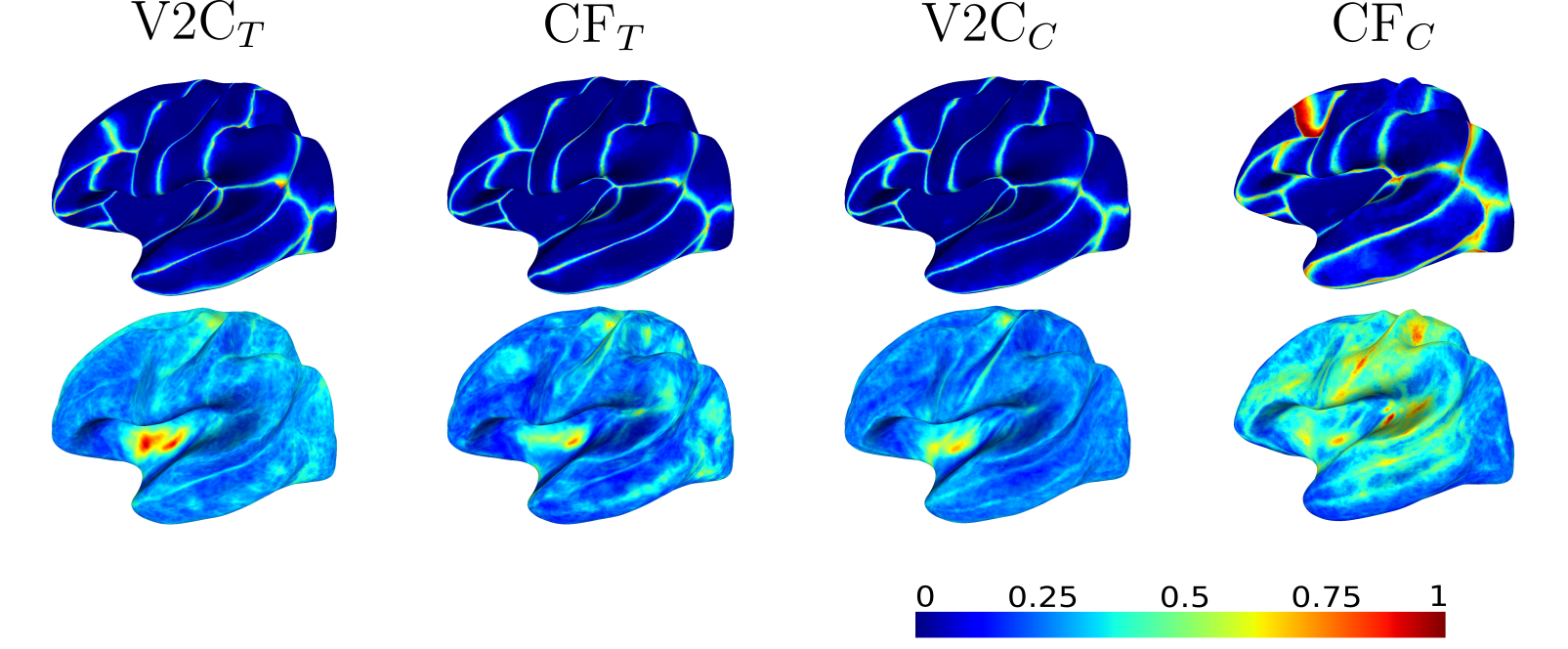}
 \caption{Visualization of parcellation and reconstruction accuracy for our four methods, averaged over the predicted left white surfaces of the OASIS test set.
 Top row: parcellation error, blue (0.0) = vertex classified correctly for all subjects, red (1.0) = vertex classified incorrectly for all subjects. Errors are mostly present in parcel boundaries.
 Bottom row: average distance from predicted surface to ground-truth surface, in mm. }
 \label{fig:results}
\end{figure}
%\end{itemize}
%\input{latex/tables/Mindboggle}
\begin{table}
\footnotesize
\setlength{\tabcolsep}{4pt} % Reduce white-space between columns, SHOULD BE SET AFTER TABLE IS FINISHED
\caption{Comparison of surface and parcellation quality of our extended Vox2Cortex (V2C) and CorticalFlow (CF) methods on the OASIS test set. Surface reconstruction metrics AD and HD are in mm. All metrics are averaged between left and right hemispheres and standard deviations are shown.}
%0.453 \interval{0.072} 
\ra{1.2}
\begin{adjustbox}{width=1\textwidth}

\begin{tabular}{@{}l c  c c @{\hskip 3\tabcolsep} c c c@{}}
    \toprule
    & \multicolumn{3}{c}{White surfaces} 
    & \multicolumn{3}{c}{Pial surfaces}
    \\
    \cmidrule(lr){2-4} \cmidrule(lr){5-7}
     & Parcellation & \multicolumn{2}{c}{Surface accuracy} & Parcellation & \multicolumn{2}{c}{Surface accuracy}
     \\
    \cmidrule(lr){2-2} \cmidrule(lr){3-4} \cmidrule(lr){5-5} \cmidrule(lr){6-7}
    Method & Dice$\uparrow$ & AD$\downarrow$ & HD$\downarrow$  & Dice$\uparrow$ & AD$\downarrow$ & HD$\downarrow$\\\midrule
   
    % baselines
    \multicolumn{7}{c}{CorticalFlow (CF)~\cite{lebrat2021corticalflow}}\\
    \midrule
    % model without parcellation, with direct mapping of fsaverage template atlas
    CF + atlas &0.810 \interval{0.095} &0.244 \interval{0.040} & 0.578 \interval{0.101} & 0.787 \interval{0.091} & 0.302 \interval{0.039}& 0.747 \interval{0.117} \\
    % model without parcellation, + Freesurfer sphere registration
    CF + FS & 0.885 \interval{0.069} &0.244 \interval{0.040} & 0.578 \interval{0.101} & n.a. & 0.302 \interval{0.039}& 0.747 \interval{0.117} \\
    $\text{CF}_C$ & 0.727 \interval{0.178} & 0.471 \interval{0.047} &1.190 \interval{0.170} & 0.672 \interval{0.178}& 0.355 \interval{0.040} & 0.896 \interval{0.126} \\
    $\text{CF}_T$ &\textbf{0.904} \interval{0.048} & 0.323 \interval{0.048}& 0.784 \interval{0.126}& \textbf{0.877} \interval{0.049}& 0.347 \interval{0.044} & 0.854 \interval{0.120}\\ %[1.2em]
    
    \midrule
     \multicolumn{7}{c}{Vox2Cortex (V2C)~\cite{Bongratz_2022_CVPR}}\\
    \midrule
    % V2C with fsaverage direct mapping of atlas
    V2C + atlas & 0.740 \interval{0.121}& 0.282 \interval{0.034}& 0.587 \interval{0.078}&0.691 \interval{0.132}& 0.341 \interval{0.037} & 0.848 \interval{0.124} \\
    % V2C + Freesurfers sphere registration 
    V2C + FS & 0.876 \interval{0.076} & 0.282 \interval{0.034}& 0.587 \interval{0.078}& n.a. & 0.341 \interval{0.037} & 0.848 \interval{0.124} \\
    $\text{V2C}_C$ & \textbf{0.902} \interval{0.050} & 0.303 \interval{0.034} & 0.641 \interval{0.082}& \textbf{0.876} \interval{0.053} & 0.362 \interval{0.038} & 0.894 \interval{0.119}\\
    $\text{V2C}_T$ & 0.885 \interval{0.057} &0.372 \interval{0.051} & 0.823 \interval{0.108}& 0.858 \interval{0.058} & 0.429 \interval{0.052}& 1.066 \interval{0.182} \\

%\hdashline
\midrule
    FastSurfer~\cite{henschel2020}  & 0.862 \interval{0.084} & 0.138 \interval{0.057} & 0.331 \interval{0.172} & 0.839 \interval{0.081} & 0.240
    \interval{0.065}& 0.557 \interval{0.179} \\

    \bottomrule
\end{tabular}
\end{adjustbox}

\label{tab:comparison2}

\vspace{-2mm}

\end{table}

\section{Results} \label{sec:results}
In the following, we show results for both of the proposed parcellation approaches, i.e., the classification network and the class-based reconstruction. To this end, we combine the proposed methods with Vox2Cortex (V2C) and CorticalFlow (CF) as described in \Cref{sec:method}. This leads to a total of four methods, which we denote as $\text{V2C}_C$, $\text{CF}_C$ (classification network) and $\text{V2C}_T$, $\text{CF}_T$ (class-based reconstruction via template). We compare our approaches to FastSurfer~\cite{henschel2020} and two additional baselines per reconstruction method. The latter are obtained by (1) training ``vanilla'' CF and V2C and mapping the atlas labels simply to the predicted surfaces (denoted as CF + atlas and V2C + atlas) and (2) using FreeSurfer's spherical atlas registration as an ad-hoc parcellation of given surfaces in a post-processing fashion (denoted as CF + FS and V2C + FS). 
The runtime for FastSurfer is about one hour, for the FS parcellation of CF and V2C meshes several hours, and for the proposed methods in the range of seconds. 
Table~\ref{tab:comparison2} presents the parcellation accuracy in terms of average Dice coefficient over all parcellation classes (computed on the surfaces). In addition, we compare the surface reconstruction accuracy in terms of average symmetric surface distance (AD) and 90th percentile Hausdorff distance (HD) in mm. 

%Baselines
We observe that FastSurfer leads to highly accurate surfaces compared to the FreeSurfer silver standard, which is expected as FastSurfer makes use of the FreeSurfer surface stream to generate surfaces. The baseline CF and V2C models also provide very accurate predictions in terms of surface accuracy with a slight advantage on the side of CF (probably due to the higher image and mesh resolution at training time). However, as expected, CF and V2C do not yield an accurate surface parcellation if a population atlas is used as their input.
Generating the DKT parcellation with FreeSurfer's atlas registration yields a higher Dice score than FastSurfer, which we attribute to the superiority of the mesh-based parcellation compared to a voxel-based approach. Note that the FreeSurfer spherical registration only works on white matter surfaces and, thus, is not applicable for the parcellation of pial surfaces. 
%and our predictions do not have vertex correspondences between white and pial surfaces.

Regarding the surface quality, we observe that solving the additional task of cortex parcellation causes a slight loss of surface accuracy in all models. This effect is most severe in the $\text{CF}_C$ and $\text{V2C}_T$ models. As the mesh-deformation network in V2C already requires several regularization losses, we suspect that the restrictive class-based reconstruction loss might interfere with the regularizers. 
In terms of parcellation accuracy, we observe best results for $\text{CF}_T$ and $\text{V2C}_C$ models with an average Dice score greater than 0.9 for white surfaces and 0.87 for pial surfaces over all parcels.
The classification GNN in $\text{V2C}_C$ takes the vertex features of the previous GNNs as input. Consequently, it can make use of vertex-wise information, which is not available in $\text{CF}_C$ (in this case, the classification network only gets vertex locations as input). As expected, $\text{CF}_C$ yields a lower parcellation accuracy compared to $\text{V2C}_C$. Therefore, we conclude that a combination of CF with a GNN classification network is not an optimal choice. 

We visualize the parcellation and surface reconstruction accuracy of the left white surfaces in Figure~\ref{fig:results} and observe that, averaged over the test set, classification errors occur almost exclusively at parcel boundaries.
Visualizations of pial surfaces are shown in the supplement.  
Overall, we conclude that the GNN classifier is better suited for V2C than for CF, as the previous graph convolutions provide more meaningful vertex input features. In contrast, the class-based reconstruction loss leads to better results in CF.

\section{Conclusion}
In this work, we introduced two effective extensions to brain reconstruction networks for joint cortex parcellation: one based on a graph classifier and one based on a novel and generic region-based reconstruction loss. Both methods are particularly suited to augment mesh-deformation networks, which provide highly accurate surface meshes, with the ability to parcellate the surfaces into associated regions. The extremely short runtime of the presented algorithms, which lies in the range of seconds at inference time, together with the high parcellation accuracy paves the way for a more fine-grained analysis of brain diseases in large-cohort studies and the integration in clinical practice.  

\paragraph*{Acknowledgments}
This research was partially supported by
the Bavarian State Ministry of Science and the Arts and coordinated by the bidt, and the BMBF (DeepMentia, 031L0200A). We gratefully acknowledge the computational resources provided by the Leibniz Supercomputing Centre (www.lrz.de).

%
%\newpage
\bibliographystyle{splncs04}
\bibliography{references}
\end{document}